\documentstyle[a4,12pt]{article}
\renewcommand{\baselinestretch}{1.35}
\def\thefootnote{\fnsymbol{footnote}}

\begin{document}
\parskip=5pt plus 1pt minus 1pt

\newpage

\begin{flushright}
{\bf DPNU Preprint} \\
{\sf hep-ex/9709005}
\end{flushright}

\vspace{0.2cm}

\begin{center}
{\Large\bf $CP$-violating Parameter $\epsilon^{~}_B$ in 
$B_d^0$-$\bar{B}^0_d$ Mixing}
\end{center}

\vspace{0.3cm}

\begin{center}
{\bf Zhi-zhong Xing} \footnote{Electronic address: xing@eken.phys.nagoya-u.ac.jp}
\end{center}

\begin{center}
{\it Department of Physics, Nagoya University, Chikusa-ku, Nagoya 464-01, Japan}
\end{center}

\vspace{3cm}

\begin{abstract}
We make an analysis of the phase-convention dependence of
the $CP$-violating parameter $\epsilon^{~}_B$ 
and give some comments on the reported CLEO and OPAL
constraints on ${\rm Re}\epsilon^{~}_B$. It is emphasized that
$\epsilon^{~}_B$ has little advantage for the study of $CP$ violation
in the $B^0_d$-$\bar{B}^0_d$ system.
\end{abstract}

\vspace{2cm}

\begin{center}
PACS number(s): 14.40.Jz, 11.30.Er, 12.15.Ff, 13.20.Jf
\end{center}

\newpage

{\Large\bf 1} ~ It has been known for more than 30 years that there exists $CP$
violation in the $K^0 \Leftrightarrow \bar{K}^0$ transition
\cite{CP64}. This effect, described commonly by a complex parameter
$\epsilon^{~}_K$, is only of order $10^{-3}$ (explicitly,
$|\epsilon^{~}_K| \approx 2.3 \times 10^{-3}$ and ${\rm
Re}\epsilon^{~}_K \approx {\rm Im} \epsilon^{~}_K$) \cite{PDG96}. So far, 
no other evidence for $CP$ violation has been unambiguously
established from a series of measurements. Some sophisticated
experimental efforts, in particular the programs of KEK and SLAC
$B$-meson factories, are underway to discover new signals of $CP$
asymmetries beyond the $K^0$-$\bar{K}^0$ system and to probe the
origin of $CP$ violation.

Analogous to $\epsilon^{~}_K$ in the neutral kaon system, a
complex parameter $\epsilon^{~}_B$ has also been used in some literature 
to describe $CP$ violation in $B^0_d$-$\bar{B}^0_d$ mixing. 
Both $\epsilon^{~}_K$ and $\epsilon^{~}_B$ are rephasing-variant
quantities, i.e., their magnitudes and phases depend on the specific
phase conventions of flavor mixings adopted for the $K^0$-$\bar{K}^0$ and
$B^0_d$-$\bar{B}^0_d$ systems. A full determination of
$\epsilon^{~}_B$, unlike the case for $\epsilon^{~}_K$, is very
difficult due to the lack of available measurements of the
$\epsilon^{~}_B$-induced $CP$ violation. 
At present the only constraint on $\epsilon^{~}_B$ is obtained from
measuring the decay-rate asymmetry between two semileptonic
channels $B^0_d \rightarrow l^+ \nu^{~}_l X^-$ and $\bar{B}^0_d
\rightarrow l^- \bar{\nu}^{~}_l X^+$, denoted by ${\cal A}_{\rm SL}$
in the following:
$$
{\cal A}_{\rm SL} \; = \; 0.031 ~ \pm ~ 0.096 ~ \pm ~ 0.032 ~~~~~~~ ({\rm CLEO
~ \cite{CLEO93}}) \; , 
$$
$$
{\cal A}_{\rm SL} \; = \; 0.008 ~ \pm ~ 0.028 ~ \pm ~ 0.012 ~~~~~~~ ({\rm OPAL
~ \cite{OPAL97}}) \; . \nonumber
$$
In the assumption of $|{\rm Im}\epsilon^{~}_B| \ll 1$, the CLEO measurement
yields $|{\rm Re} \epsilon^{~}_B| < 0.045$ at the $90\%$ confidence
level, and the OPAL measurement announces ${\rm Re}\epsilon^{~}_B =
0.002 \pm 0.009 \pm 0.003$ with a higher degree of accuracy.

In this short note we shall discuss the phase-convention dependence of 
$\epsilon^{~}_B$ and comment on the CLEO 
and OPAL constraints on ${\rm Re}\epsilon^{~}_B$.
Within the standard model we emphasize that $|{\rm Im}\epsilon^{~}_B| \ll 1$
taken in the CLEO and OPAL analyses corresponds to an unusual phase
convention of the Cabibbo-Kobayashi-Maskawa (CKM) matrix $V$, in which
$V_{tb}$ and $V_{td}$ are essentially real. If the ``standard''
parametrization of the CKM
matrix is adopted, however, one will arrive at $|\epsilon^{~}_B|
\approx {\rm Im}\epsilon^{~}_B \approx \tan\beta$, where $\beta$ is 
an angle of the well-known CKM unitarity triangle. In
this case we obtain $0.16 \leq |\epsilon^{~}_B| \leq 0.70$ by use of 
current data, hence the $|\epsilon^{~}_B|^2$ correction to
$CP$-violating signals of $B_d$ decays cannot always be neglected.
It is concluded that the rephasing-variant parameter $\epsilon^{~}_B$
has little advantage for the study of $CP$ violation 
in the $B^0_d$-$\bar{B}^0_d$ system.

{\Large\bf 2} ~ In the assumption of $CPT$ invariance, the mass eigenstates of $B^0_d$ 
and $\bar{B}^0_d$ mesons can be written as 
\begin{eqnarray}
|B_1 \rangle & = & p^{~}_B |B^0_d\rangle ~ + ~ q^{~}_B |\bar{B}^0_d\rangle \; 
, \nonumber \\
|B_2 \rangle & = & p^{~}_B |B^0_d\rangle ~ - ~ q^{~}_B |\bar{B}^0_d \rangle
\; .
\end{eqnarray}
where $p^{~}_B$ and $q^{~}_B$ are complex mixing parameters with the normalization 
$|p^{~}_B|^2 + |q^{~}_B|^2 = 1$. In terms of the $\epsilon^{~}_B$ parameter, $p^{~}_B$
and $q^{~}_B$ read
\begin{equation}
p^{~}_B \; =\; \frac{1 + \epsilon^{~}_B}{\sqrt{2 (1 +
|\epsilon^{~}_B|^2)}} \; , ~~~~~~~~ 
q^{~}_B \; =\; \frac{1 - \epsilon^{~}_B}{\sqrt{2 (1 +
|\epsilon^{~}_B|^2)}} \; , ~~~~~~~~ 
\end{equation}
Note that both $p^{~}_B$ and $q^{~}_B$ are dependent on the phase
conventions for $B^0_d$ and $\bar{B}^0_d$ states which are defined by
flavor-conserving strong interactions \cite{Tsai96}. Only $|q^{~}_B/p^{~}_B| =
|(1-\epsilon^{~}_B)/(1+\epsilon^{~}_B)|$ is rephasing-inviarant.

It is known that the $CP$ asymmetry between the wrong-sign events of
incoherent semileptonic $B_d$ decays is identical to that between the
same-sign dilepton events of coherent $B^0_d\bar{B}^0_d$ decays at the 
$\Upsilon (4S)$ resonance \cite{BigiSanda81}. 
This asymmetry can be given as follows:
\begin{equation}
{\cal A}_{\rm SL} \; =\; \frac{|p^{~}_B|^4 - |q^{~}_B|^4}{|p^{~}_B|^4 + 
|q^{~}_B|^4} \; =\; \frac{4 {\rm Re}\epsilon^{~}_B \left ( 1 +
|\epsilon^{~}_B|^2 \right )}{ \left (1 + |\epsilon^{~}_B|^2 \right )^2 + 4 ({\rm
Re} \epsilon^{~}_B )^2} \; .
\end{equation}
The standard model predicts $|{\cal A}_{\rm SL}| \sim
10^{-3}$ (see, e.g., Ref. \cite{SM}), but the
presence of new physics in $B^0_d$-$\bar{B}^0_d$ mixing might
enhance the magnitude of ${\cal A}_{\rm SL}$ up to the percent level
\cite{Maiani92,SandaXing97}.

\begin{figure}
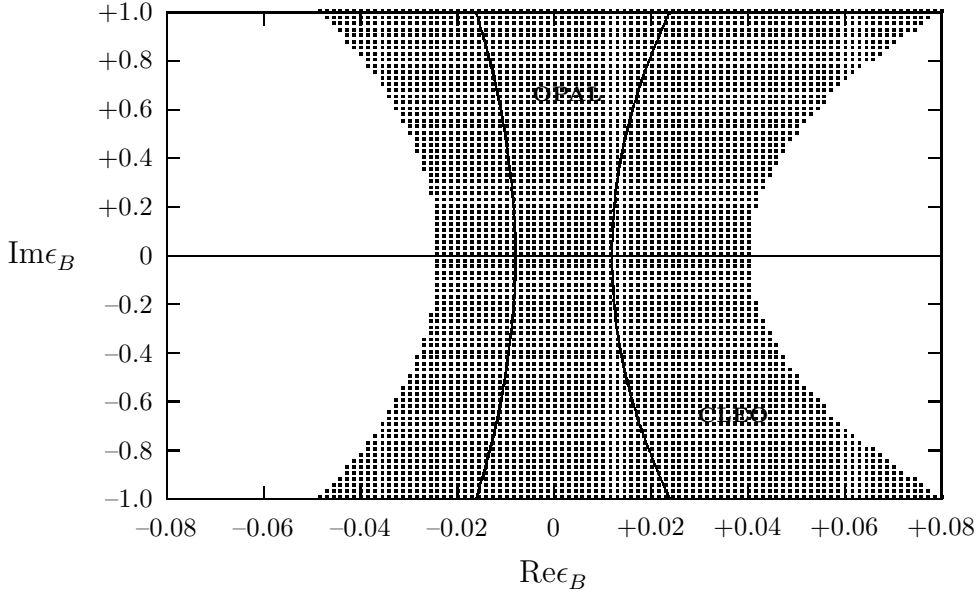

\setlength{\unitlength}{0.240900pt}
\ifx\plotpoint\undefined\newsavebox{\plotpoint}\fi
\sbox{\plotpoint}{\rule[-0.200pt]{0.400pt}{0.400pt}}%

\caption{\small The $({\rm Re}\epsilon^{~}_B, {\rm Im}\epsilon^{~}_B)$
plot constrained by CLEO (the whole dotted region) and OPAL 
(the dotted region between two solid curves) measurements of the
lepton $CP$ asymmetry ${\cal A}_{\rm SL}$.}
\end{figure}
The CLEO and OPAL measurements of ${\cal A}_{\rm SL}$ impose a
constraint on $\epsilon^{~}_B$, which can be illustrated by a $({\rm
Re}\epsilon^{~}_B, {\rm Im}\epsilon^{~}_B)$ plot. Conservatively we
scan the measured value of ${\cal A}_{\rm SL}$ within its error bar, and
depict the allowed region for ${\rm Re}\epsilon^{~}_B$ and ${\rm
Im}\epsilon^{~}_B$ in Fig. 1, where ${\rm Im}\epsilon^{~}_B \in [-1, + 
1]$ has been taken.
It becomes clear that CLEO and OPAL constraints on ${\rm
Re}\epsilon^{~}_B$ result from the phase convention ${\rm
Im}\epsilon^{~}_B =0$ or $|{\rm Im}\epsilon^{~}_B| \ll 1$ \cite{Nir97}. 
If $|{\rm Im}\epsilon^{~}_B| \sim O(1)$ is taken, however, 
a new upper bound $|{\rm Re}\epsilon^{~}_B| < 0.1$ may appear as a
straightforward consequence of the CLEO measurement of ${\cal A}_{\rm SL}$.

In the standard model, the box-diagram calculation of
$B^0_d$-$\bar{B}^0_d$ mixing yields
\begin{equation}
\frac{q^{~}_B}{p^{~}_B} \; \approx \; \frac{ 1 - i {\rm Im} \epsilon^{~}_B }{ 1 + 
i {\rm Im} \epsilon^{~}_B } \; \approx \; \frac{V_{td}
V^*_{tb}}{V_{td}^* V_{tb}} \; \equiv \; e^{-2i\phi_M} \; ,
\end{equation}
where $\phi_M$ denotes the mixing phase. Hence the phase convention
$|{\rm Im}\epsilon^{~}_B| \ll 1$ taken above 
implies $\phi_M \approx 0$. In fact this
corresponds to a specific parametrization of the CKM
matrix, such as the following form \cite{FritzschXing97}:
\begin{equation}
V \; =\; \left ( \matrix{
s_{\rm u} s_{\rm d} c + c_{\rm u} c_{\rm d} e^{-{\rm i}\varphi} &
s_{\rm u} c_{\rm d} c - c_{\rm u} s_{\rm d} e^{-{\rm i}\varphi} &
s_{\rm u} s \cr
c_{\rm u} s_{\rm d} c - s_{\rm u} c_{\rm d} e^{-{\rm i}\varphi} &
c_{\rm u} c_{\rm d} c + s_{\rm u} s_{\rm d} e^{-{\rm i}\varphi}   &
c_{\rm u} s \cr
- s_{\rm d} s	& - c_{\rm d} s	& c \cr } \right ) \; ,
\end{equation}
where $s \equiv \sin\theta$, $s_{\rm u} \equiv \sin\theta_{\rm u}$,
etc. In contrast, the ``standard'' parametrization \cite{PDG96} (or 
the Wolfenstein parametrization \cite{Wolfenstein83})
cannot accommodate the phase convention $\phi_M =0$.

{\Large\bf 3} ~ Next we give a further illustration of the phase-convention dependence
of CLEO and OPAL results by estimating the magnitude of
${\rm Im} \epsilon^{~}_B$ in the ``standard'' parametrization \cite{PDG96}:
\begin{equation}
V \; =\; \left ( \matrix{
c_{12} c_{13}	& s_{12} c_{13} 	& s_{13} e^{-i\delta_{13}} \cr 
-s_{12} c_{23} - c_{12} s_{23} s_{13} e^{i\delta_{13}}	& c_{12}
c_{23} - s_{12} s_{23} s_{13} e^{i\delta_{13}}	& s_{23} c_{13} \cr
s_{12} s_{23} - c_{12} c_{23} s_{13} e^{i\delta_{13}}	& -c_{12}
s_{23} - s_{12} c_{23} s_{13} e^{i\delta_{13}}	& c_{23} c_{13} \cr}
\right ) \; ,
\end{equation}
where $s_{12} \equiv \sin\theta_{12}$, etc.
To be specific, we assume $|{\rm Im}\epsilon^{~}_B| \leq 1$; then 
$|{\rm Re}\epsilon^{~}_B| < 0.1$ holds (as shown in Fig. 1) and its
contribution to $q^{~}_B/p^{~}_B$ can be neglected in the lowest order
approximation. 
In this case, we obtain 
\begin{equation}
\tan (2\phi_M) \; \approx \; \frac{2 {\rm Im}\epsilon^{~}_B}{ 1 -
({\rm Im} \epsilon^{~}_B )^2} \; 
\end{equation}
from Eq. (4); i.e., ${\rm Im} \epsilon^{~}_B \approx \tan\phi_M$ as a
function of $\delta_{13}$. Furthermore, we have 
$\phi_M \approx \beta$ in the present parametrization, 
where $\beta$ is an inner angle of the 
well-known CKM unitarity triangle \cite{PDG96}. A careful analysis of current data
on $\epsilon^{~}_K$, $|V_{ub}/V_{cb}|$ and $B^0_d$-$\bar{B}^0_d$
mixing has yielded $0.32 \leq \sin (2\beta) \leq 0.94$ at the $95\%$
confidence level \cite{Ali96}. Therefore we get
\begin{equation}
0.16 \; \leq \; {\rm Im} \epsilon^{~}_B \; \leq \; 0.70 \; .
\end{equation}
This result is consistent with our assumption made above (i.e., $|{\rm Im}\epsilon^{~}_B| 
\leq 1$).

The magnitude of $\epsilon^{~}_B$ is mainly governed by that of ${\rm
Im}\epsilon^{~}_B$. If we take the OPAL measurement seriously, it
turns out that ${\rm Re}\epsilon^{~}_B <2\%$ in the present phase
convention. In this case, $|\epsilon^{~}_B| \approx {\rm
Im}\epsilon^{~}_B$ holds approximately. Thus the $|\epsilon^{~}_B|^2$ correction to $CP$ 
asymmetries of weak $B_d$ decays, usually appearing in the
form of a dilution factor 
\begin{equation}
F_{\rm D} \; \equiv \; \frac{1}{1 + |\epsilon^{~}_B|^2} \; ,
\end{equation}
cannot be neglected in some cases. The behavior of $F_{\rm D}$ changing with
$|\epsilon^{~}_B|$ is illustrated in Fig. 2.
We obtain $0.67 \leq F_{\rm D} \leq 0.98$,
corresponding to the allowed range of ${\rm Im}\epsilon^{~}_B$ in Eq. (8).
The lepton $CP$ aymmetry given in Eq. (3) can be
approximated to 
\begin{equation}
{\cal A}_{\rm SL} \; \approx \; 4 ~ F_{\rm D} ~ {\rm Re} \epsilon^{~}_B \; 
\approx \; (2.7 \sim 3.9) \times {\rm Re} \epsilon^{~}_B \; , 
\end{equation}
due to the smallness of $|{\rm Re}\epsilon^{~}_B|$. 
Thus the magnitude of ${\rm Re}\epsilon^{~}_B$ obtained here
is about $1/F_{\rm D} \approx 1.0 \sim 1.5$ times
larger than that reported by OPAL \cite{OPAL97} with the convention
$|{\rm Im}\epsilon^{~}_B| \ll 1$.
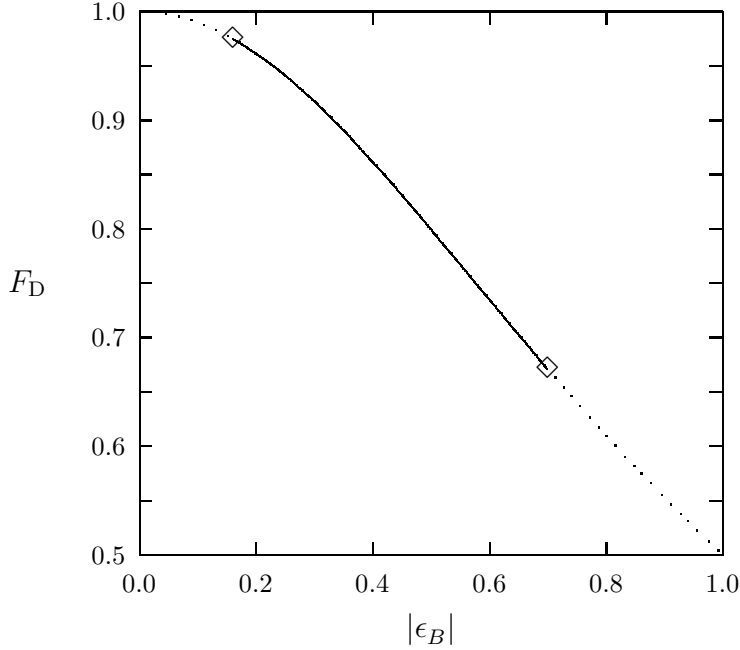
\begin{figure}
\setlength{\unitlength}{0.240900pt}
\ifx\plotpoint\undefined\newsavebox{\plotpoint}\fi
\sbox{\plotpoint}{\rule[-0.200pt]{0.400pt}{0.400pt}}%
\begin{picture}(1200,990)(-350,0)
\font\gnuplot=cmr10 at 10pt
\gnuplot
\sbox{\plotpoint}{\rule[-0.200pt]{0.400pt}{0.400pt}}%
\put(220.0,113.0){\rule[-0.200pt]{0.400pt}{205.729pt}}
\put(220.0,113.0){\rule[-0.200pt]{4.818pt}{0.400pt}}
\put(198,113){\makebox(0,0)[r]{0.5}}
\put(1116.0,113.0){\rule[-0.200pt]{4.818pt}{0.400pt}}
\put(220.0,198.0){\rule[-0.200pt]{4.818pt}{0.400pt}}
\put(1116.0,198.0){\rule[-0.200pt]{4.818pt}{0.400pt}}
\put(220.0,284.0){\rule[-0.200pt]{4.818pt}{0.400pt}}
\put(198,284){\makebox(0,0)[r]{0.6}}
\put(1116.0,284.0){\rule[-0.200pt]{4.818pt}{0.400pt}}
\put(220.0,369.0){\rule[-0.200pt]{4.818pt}{0.400pt}}
\put(1116.0,369.0){\rule[-0.200pt]{4.818pt}{0.400pt}}
\put(220.0,455.0){\rule[-0.200pt]{4.818pt}{0.400pt}}
\put(198,455){\makebox(0,0)[r]{0.7}}
\put(1116.0,455.0){\rule[-0.200pt]{4.818pt}{0.400pt}}
\put(220.0,540.0){\rule[-0.200pt]{4.818pt}{0.400pt}}
\put(1116.0,540.0){\rule[-0.200pt]{4.818pt}{0.400pt}}
\put(220.0,625.0){\rule[-0.200pt]{4.818pt}{0.400pt}}
\put(198,625){\makebox(0,0)[r]{0.8}}
\put(1116.0,625.0){\rule[-0.200pt]{4.818pt}{0.400pt}}
\put(220.0,711.0){\rule[-0.200pt]{4.818pt}{0.400pt}}
\put(1116.0,711.0){\rule[-0.200pt]{4.818pt}{0.400pt}}
\put(220.0,796.0){\rule[-0.200pt]{4.818pt}{0.400pt}}
\put(198,796){\makebox(0,0)[r]{0.9}}
\put(1116.0,796.0){\rule[-0.200pt]{4.818pt}{0.400pt}}
\put(220.0,882.0){\rule[-0.200pt]{4.818pt}{0.400pt}}
\put(1116.0,882.0){\rule[-0.200pt]{4.818pt}{0.400pt}}
\put(220.0,967.0){\rule[-0.200pt]{4.818pt}{0.400pt}}
\put(198,967){\makebox(0,0)[r]{1.0}}
\put(1116.0,967.0){\rule[-0.200pt]{4.818pt}{0.400pt}}
\put(220.0,113.0){\rule[-0.200pt]{0.400pt}{4.818pt}}
\put(220,68){\makebox(0,0){0.0}}
\put(220.0,947.0){\rule[-0.200pt]{0.400pt}{4.818pt}}
\put(403.0,113.0){\rule[-0.200pt]{0.400pt}{4.818pt}}
\put(403,68){\makebox(0,0){0.2}}
\put(403.0,947.0){\rule[-0.200pt]{0.400pt}{4.818pt}}
\put(586.0,113.0){\rule[-0.200pt]{0.400pt}{4.818pt}}
\put(586,68){\makebox(0,0){0.4}}
\put(586.0,947.0){\rule[-0.200pt]{0.400pt}{4.818pt}}
\put(770.0,113.0){\rule[-0.200pt]{0.400pt}{4.818pt}}
\put(770,68){\makebox(0,0){0.6}}
\put(770.0,947.0){\rule[-0.200pt]{0.400pt}{4.818pt}}
\put(953.0,113.0){\rule[-0.200pt]{0.400pt}{4.818pt}}
\put(953,68){\makebox(0,0){0.8}}
\put(953.0,947.0){\rule[-0.200pt]{0.400pt}{4.818pt}}
\put(1136.0,113.0){\rule[-0.200pt]{0.400pt}{4.818pt}}
\put(1136,68){\makebox(0,0){1.0}}
\put(1136.0,947.0){\rule[-0.200pt]{0.400pt}{4.818pt}}
\put(220.0,113.0){\rule[-0.200pt]{220.664pt}{0.400pt}}
\put(1136.0,113.0){\rule[-0.200pt]{0.400pt}{205.729pt}}
\put(220.0,967.0){\rule[-0.200pt]{220.664pt}{0.400pt}}
\put(45,540){\makebox(0,0){$F_{\rm D}$}}
\put(678,-3){\makebox(0,0){$|\epsilon^{~}_B|$}}
\put(220.0,113.0){\rule[-0.200pt]{0.400pt}{205.729pt}}
\put(367,924){\usebox{\plotpoint}}
\multiput(367.00,922.92)(0.826,-0.492){19}{\rule{0.755pt}{0.118pt}}
\multiput(367.00,923.17)(16.434,-11.000){2}{\rule{0.377pt}{0.400pt}}
\multiput(385.00,911.92)(0.755,-0.492){21}{\rule{0.700pt}{0.119pt}}
\multiput(385.00,912.17)(16.547,-12.000){2}{\rule{0.350pt}{0.400pt}}
\multiput(403.00,899.92)(0.734,-0.493){23}{\rule{0.685pt}{0.119pt}}
\multiput(403.00,900.17)(17.579,-13.000){2}{\rule{0.342pt}{0.400pt}}
\multiput(422.00,886.92)(0.644,-0.494){25}{\rule{0.614pt}{0.119pt}}
\multiput(422.00,887.17)(16.725,-14.000){2}{\rule{0.307pt}{0.400pt}}
\multiput(440.00,872.92)(0.600,-0.494){27}{\rule{0.580pt}{0.119pt}}
\multiput(440.00,873.17)(16.796,-15.000){2}{\rule{0.290pt}{0.400pt}}
\multiput(458.00,857.92)(0.561,-0.494){29}{\rule{0.550pt}{0.119pt}}
\multiput(458.00,858.17)(16.858,-16.000){2}{\rule{0.275pt}{0.400pt}}
\multiput(476.00,841.92)(0.558,-0.495){31}{\rule{0.547pt}{0.119pt}}
\multiput(476.00,842.17)(17.865,-17.000){2}{\rule{0.274pt}{0.400pt}}
\multiput(495.00,824.92)(0.498,-0.495){33}{\rule{0.500pt}{0.119pt}}
\multiput(495.00,825.17)(16.962,-18.000){2}{\rule{0.250pt}{0.400pt}}
\multiput(513.00,806.92)(0.498,-0.495){33}{\rule{0.500pt}{0.119pt}}
\multiput(513.00,807.17)(16.962,-18.000){2}{\rule{0.250pt}{0.400pt}}
\multiput(531.00,788.92)(0.498,-0.495){35}{\rule{0.500pt}{0.119pt}}
\multiput(531.00,789.17)(17.962,-19.000){2}{\rule{0.250pt}{0.400pt}}
\multiput(550.58,768.74)(0.495,-0.554){33}{\rule{0.119pt}{0.544pt}}
\multiput(549.17,769.87)(18.000,-18.870){2}{\rule{0.400pt}{0.272pt}}
\multiput(568.58,748.74)(0.495,-0.554){33}{\rule{0.119pt}{0.544pt}}
\multiput(567.17,749.87)(18.000,-18.870){2}{\rule{0.400pt}{0.272pt}}
\multiput(586.58,728.84)(0.495,-0.525){35}{\rule{0.119pt}{0.521pt}}
\multiput(585.17,729.92)(19.000,-18.919){2}{\rule{0.400pt}{0.261pt}}
\multiput(605.58,708.65)(0.495,-0.583){33}{\rule{0.119pt}{0.567pt}}
\multiput(604.17,709.82)(18.000,-19.824){2}{\rule{0.400pt}{0.283pt}}
\multiput(623.58,687.65)(0.495,-0.583){33}{\rule{0.119pt}{0.567pt}}
\multiput(622.17,688.82)(18.000,-19.824){2}{\rule{0.400pt}{0.283pt}}
\multiput(641.58,666.66)(0.495,-0.578){35}{\rule{0.119pt}{0.563pt}}
\multiput(640.17,667.83)(19.000,-20.831){2}{\rule{0.400pt}{0.282pt}}
\multiput(660.58,644.56)(0.495,-0.611){33}{\rule{0.119pt}{0.589pt}}
\multiput(659.17,645.78)(18.000,-20.778){2}{\rule{0.400pt}{0.294pt}}
\multiput(678.58,622.56)(0.495,-0.611){33}{\rule{0.119pt}{0.589pt}}
\multiput(677.17,623.78)(18.000,-20.778){2}{\rule{0.400pt}{0.294pt}}
\multiput(696.58,600.66)(0.495,-0.578){35}{\rule{0.119pt}{0.563pt}}
\multiput(695.17,601.83)(19.000,-20.831){2}{\rule{0.400pt}{0.282pt}}
\multiput(715.58,578.56)(0.495,-0.611){33}{\rule{0.119pt}{0.589pt}}
\multiput(714.17,579.78)(18.000,-20.778){2}{\rule{0.400pt}{0.294pt}}
\multiput(733.58,556.56)(0.495,-0.611){33}{\rule{0.119pt}{0.589pt}}
\multiput(732.17,557.78)(18.000,-20.778){2}{\rule{0.400pt}{0.294pt}}
\multiput(751.58,534.66)(0.495,-0.578){35}{\rule{0.119pt}{0.563pt}}
\multiput(750.17,535.83)(19.000,-20.831){2}{\rule{0.400pt}{0.282pt}}
\multiput(770.58,512.56)(0.495,-0.611){33}{\rule{0.119pt}{0.589pt}}
\multiput(769.17,513.78)(18.000,-20.778){2}{\rule{0.400pt}{0.294pt}}
\multiput(788.58,490.56)(0.495,-0.611){33}{\rule{0.119pt}{0.589pt}}
\multiput(787.17,491.78)(18.000,-20.778){2}{\rule{0.400pt}{0.294pt}}
\multiput(806.58,468.66)(0.495,-0.578){35}{\rule{0.119pt}{0.563pt}}
\multiput(805.17,469.83)(19.000,-20.831){2}{\rule{0.400pt}{0.282pt}}
\multiput(825.58,446.56)(0.495,-0.611){33}{\rule{0.119pt}{0.589pt}}
\multiput(824.17,447.78)(18.000,-20.778){2}{\rule{0.400pt}{0.294pt}}
\multiput(843.58,424.56)(0.495,-0.611){33}{\rule{0.119pt}{0.589pt}}
\multiput(842.17,425.78)(18.000,-20.778){2}{\rule{0.400pt}{0.294pt}}
\put(220,967){\usebox{\plotpoint}}
\put(220.00,967.00){\usebox{\plotpoint}}
\put(240.71,965.71){\usebox{\plotpoint}}
\put(261.32,963.28){\usebox{\plotpoint}}
\put(281.63,959.16){\usebox{\plotpoint}}
\put(301.54,953.30){\usebox{\plotpoint}}
\put(321.12,946.45){\usebox{\plotpoint}}
\put(340.05,937.98){\usebox{\plotpoint}}
\put(358.50,928.47){\usebox{\plotpoint}}
\put(376.51,918.19){\usebox{\plotpoint}}
\put(393.99,907.00){\usebox{\plotpoint}}
\put(411.20,895.39){\usebox{\plotpoint}}
\put(428.05,883.29){\usebox{\plotpoint}}
\put(444.31,870.40){\usebox{\plotpoint}}
\multiput(458,859)(15.513,-13.789){2}{\usebox{\plotpoint}}
\put(491.18,829.42){\usebox{\plotpoint}}
\put(506.05,814.95){\usebox{\plotpoint}}
\put(520.73,800.27){\usebox{\plotpoint}}
\put(535.40,785.60){\usebox{\plotpoint}}
\multiput(550,771)(13.885,-15.427){2}{\usebox{\plotpoint}}
\put(577.85,740.06){\usebox{\plotpoint}}
\put(591.90,724.79){\usebox{\plotpoint}}
\multiput(605,711)(13.508,-15.759){2}{\usebox{\plotpoint}}
\put(633.14,678.16){\usebox{\plotpoint}}
\put(646.68,662.43){\usebox{\plotpoint}}
\multiput(660,647)(13.143,-16.064){2}{\usebox{\plotpoint}}
\put(686.52,614.58){\usebox{\plotpoint}}
\multiput(696,603)(13.566,-15.708){2}{\usebox{\plotpoint}}
\put(726.54,566.89){\usebox{\plotpoint}}
\put(739.69,550.83){\usebox{\plotpoint}}
\multiput(751,537)(13.566,-15.708){2}{\usebox{\plotpoint}}
\put(779.71,503.13){\usebox{\plotpoint}}
\multiput(788,493)(13.143,-16.064){2}{\usebox{\plotpoint}}
\put(819.56,455.30){\usebox{\plotpoint}}
\put(832.87,439.38){\usebox{\plotpoint}}
\multiput(843,427)(13.143,-16.064){2}{\usebox{\plotpoint}}
\put(872.98,391.76){\usebox{\plotpoint}}
\put(886.69,376.19){\usebox{\plotpoint}}
\multiput(898,363)(13.508,-15.759){2}{\usebox{\plotpoint}}
\put(927.22,328.91){\usebox{\plotpoint}}
\put(940.93,313.34){\usebox{\plotpoint}}
\multiput(953,300)(13.885,-15.427){2}{\usebox{\plotpoint}}
\put(982.62,267.09){\usebox{\plotpoint}}
\put(996.93,252.07){\usebox{\plotpoint}}
\multiput(1008,241)(14.274,-15.068){2}{\usebox{\plotpoint}}
\put(1040.06,207.16){\usebox{\plotpoint}}
\put(1054.63,192.37){\usebox{\plotpoint}}
\put(1069.30,177.70){\usebox{\plotpoint}}
\multiput(1081,166)(14.676,-14.676){2}{\usebox{\plotpoint}}
\put(1113.71,134.06){\usebox{\plotpoint}}
\put(1128.80,119.80){\usebox{\plotpoint}}
\put(1136,113){\usebox{\plotpoint}}
\sbox{\plotpoint}{\rule[-0.400pt]{0.800pt}{0.800pt}}%
\put(367,924){\raisebox{-.8pt}{\makebox(0,0){$\Diamond$}}}
\put(861,405){\raisebox{-.8pt}{\makebox(0,0){$\Diamond$}}}
\end{picture}
\caption{\small The dilution factor of $|\epsilon^{~}_B|$ in the ``standard''
parametrization of the CKM matrix.}
\end{figure}

Another example is the time-integrated $CP$ asymmetry in $B_d \rightarrow J/\psi 
K_S$ on the $\Upsilon (4S)$ resonance. This signal, denoted by ${\cal 
A}_{\psi K}$ below, generally consists of three contributions:
the asymmetry arising from $B^0_d$-$\bar{B}^0_d$ mixing (${\rm Re}
\epsilon^{~}_B$), that from $K^0$-$\bar{K}^0$ mixing (${\rm Re}
\epsilon^{~}_K$), and that from the direct decay amplitude
\cite{Xing95}.
For simplicity, here we neglect the third
contribution to ${\cal A}_{\psi K}$ because its magnitude is
difficult to be evaluated in a reliable way \cite{DuXing93}. If the CKM matrix
takes the phase convention as Eq. (6), then ${\cal A}_{\psi K}$ can be 
given as follows \cite{Xing95}:
\begin{eqnarray}
{\cal A}_{\psi K} & \approx & \frac{2}{1 +x^2_d} {\rm
Re}\epsilon^{~}_K ~ - ~ \frac{2 x^2_d F_{\rm D}}{1+x^2_d} {\rm
Re}\epsilon^{~}_B \; \nonumber \\
& \approx & 2.1 \times 10^{-3} ~ - ~ (0.46 \sim 0.68) \times {\rm Re}
\epsilon^{~}_B \; ,
\end{eqnarray}
where ${\rm Re}\epsilon^{~}_K \approx 1.65\times 10^{-3}$, and $x_d
\approx 0.73$ is the $B^0_d$-$\bar{B}^0_d$ mixing parameter
\cite{PDG96}. We observe that the magnitude of ${\cal
A}_{\psi K}$ is sensitive to the dilution factor $F_{\rm
D}$. It should be more careful to deal with the
$F_{\rm D}$ correction in some $CP$ asymmetries of $B_d$ decays, if one wants to
use them to test $CPT$ symmetry in $B^0_d$-$\bar{B}^0_d$ mixing
\cite{OPAL97,CPT}.

{\Large\bf 4} ~ We have made a detailed analysis of the
phase-convention dependence of $\epsilon^{~}_B$ and given some
critical comments on the reported CLEO and OPAL constraints on ${\rm
Re}\epsilon^{~}_B$. Unlike $\epsilon^{~}_K$ in the neutral kaon
system, $\epsilon^{~}_B$ (both its magnitude and phase information)
cannot be fully determined from the available measurement of $CP$
violation associated with $B^0_d$-$\bar{B}^0_d$ mixing. Hence a given
constraint on ${\rm Re}\epsilon^{~}_B$ just reflects a special
convention for ${\rm Im}\epsilon^{~}_B$ or the CKM matrix $V$. In this 
sense, we emphasize that the rephasing-variant parameter 
$\epsilon^{~}_B$ has little advantage for the study of $CP$ violation
in weak $B_d$ decays.

{\it Acknowledgments:} ~
The author would like to thank Prof. Y. Nir for his critical comments
and useful suggestions on the first version of this note. 
Several helpful conversations with Prof. S.Y. Tsai is also
acknowledged.
Finally he is grateful to Prof. A.I. Sanda for his warm hospitality and to the
Japan Society for the Promotion of Science for its financial support.

\end{document}